\documentclass[aps,preprint,groupedaddress,superscriptaddress,longbibliography]{revtex4}

\usepackage{graphicx}
\usepackage{dcolumn}
\usepackage{bm}
\usepackage{multirow}
\usepackage{color}

\begin{document}

\title{Stacking polymorphism of PtSe$_2$: Its implication to layer-dependent metal-insulator transitions
}

\author{Jeonghwan Ahn}
\affiliation{Materials Science and Technology Division, Oak Ridge National Laboratory, Oak Ridge, Tennessee 37831, USA}
\affiliation{Department of Physics, Konkuk University, Seoul 05029, Korea}
\author{Iuegyun Hong}
\affiliation{Department of Physics, Konkuk University, Seoul 05029, Korea}
\author{Gwangyoung Lee}
\affiliation{Department of Physics, Konkuk University, Seoul 05029, Korea}
\author{Hyeondeok Shin}
\affiliation{Computational Science Division, Argonne National Laboratory, Argonne, Illinois 60439, USA}
\author{Anouar Benali}
\affiliation{Computational Science Division, Argonne National Laboratory, Argonne, Illinois 60439, USA}
\author{Yongkyung Kwon}
\email{ykwon@konkuk.ac.kr}
\affiliation{Department of Physics, Konkuk University, Seoul 05029, Korea}
\author{Jaron T. Krogel}
\email{krogeljt@ornl.gov}
\affiliation{Materials Science and Technology Division, Oak Ridge National Laboratory, Oak Ridge, Tennessee 37831, USA}

\date{\today}

\begin{abstract}
{
Using diffusion Monte Carlo (DMC) and density functional theory (DFT) calculations, we examined the structural stability and interlayer binding properties of PtSe$_2$, a representative transition metal dichalcogenide (TMD) with strong interlayer interaction. Our DMC results for the bilayer revealed that AA and AB-r stacking modes are nearly degenerate, highlighting the significant role of interlayer hybridization in offsetting the energy cost due to larger interlayer separations in the AB-r mode. Additionally, our DMC-benchmarked DFT studies with the r$^2$SCAN+rVV10 functional demonstrated pronounced stacking polymorphism in few-layer PtSe$_2$, suggesting the potential for stacking faults and the formation of grain boundaries between different stacking domains which could develop metallic electronic structures.
Thus this polymorphism, along with selenium vacancies, influences a layer-dependent metal-insulator transition observed in few-layer PtSe$_2$. Our findings emphasize the importance of both van der Waals interactions and interlayer hybridization in determining the phase stability and electronic properties of TMDs, advancing our understanding of their fundamental properties and refining theoretical models for practical applications in nanoelectronic devices.

}
\end{abstract}

\maketitle

\section*{Introduction}
\label{sec:intro}
Interlayer interactions in transition metal di-chalcogenides (TMDs) are complex, driven by both dispersive van der Waals (vdW) forces and orbital hybridization between adjacent chalcogen layers.
Unlike the non-directional vdW interactions, interlayer hybridization is significantly dependent on the hybridization axis, which is highly sensitive to the stacking configuration.
In this respect, interlayer hybridization in TMDs is expected to be more prominent when the hybridization axis aligns with the stacking axis to preferentially facilitate interactions between p$_{z}$ states of chalcogen atoms in adjacent layers.
This hybridization may cause a large splitting between anti-bonding and bonding states near the Fermi level, leading to a significant change in the band structure when forming a layered structure - one of the characteristic features of TMD materials with strong interlayer coupling.
This suggests that interlayer hybridization could be as influential as the vdW interaction in determining the phase stability and electronic properties of TMD materials.

In this work, we focus on PtSe$_2$, a representative TMD material known for its strong interlayer coupling due to pronounced Se-p$_{z}$ interlayer hybridization~\cite{zhao2017high}.
Notably, its band gap was observed to be strongly dependent on layer thickness, leading to a metal-insulator transition (MIT) as additional layers are added~\cite{yan2017high,ciarrocchi2018thickness,zhang2021precise,li2021layer}.
The strong interlayer coupling in PtSe$_2$ was also evident in various optical measurements such as Raman spectra~\cite{tharrault2024raman} and reflective index~\cite{tharrault2023optical}, which vary with the number of layers.
Interestingly, a series of these measurements have revealed various stacking modes beyond the ground-state AA stacking, each characterized by distinct optical signals.
The observation of different stacking modes led to Kempt {\it et~al.}'s suggestion of a possible stacking polymorphism in bulk PtSe$_{2}$~\cite{kempt2022stacking},
which also aligns with the frequent formation of a polycrystalline phase upon its synthesis~\cite{ansari2019quantum,jiang2019large,tharrault2024raman}. 
Furthermore, there was an experimental report that grain boundary formation between two different stacking domains in the bilayer structure can contribute to metallic transport 
through channels created by extra Pt or Se atoms in the boundary~\cite{chen2021atomic},
which is understood to be a potential source of differing experimental estimates for the critical layer count necessary for the onset of MIT~\cite{yan2017high,ciarrocchi2018thickness,zhang2021precise,li2021layer}. 
Thus, accurately identifying the onset of MIT in well-controlled PtSe$_2$ samples, free from grain boundaries and other defects, requires a thorough understanding of the relative energetics among different stacking configurations.

The theoretical description of the interlayer interactions in PtSe$_2$ is challenging because of the intricate interplay of vdW forces, interlayer hybridization, and potential metallization.
Currently available vdW-DFT functionals have produced widely scattered results for interlayer separations of layered PtSe$_{2}$ systems.
Accordingly, the DFT estimations for band gaps and the critical layer thickness for the MIT have also varied depending on the density functional employed~\cite{zhang2017mechanism,kandemir2018structural,fang2019structural,zhang2021precise}. 
Even on the experimental side, there has been considerable variations in the measurements of interlayer separation in bilayer PtSe$_{2}$, ranging from 5.0~\AA~to 5.9~\AA~\cite{shawkat2020thickness,li2021layer,zhang2021precise}. 
On the other hand, the diffusion quantum Monte Carlo (DMC) method has been successfully applied to accurately assess interlayer interactions of various 2D layered materials~\cite{mostaani15,shulenburger15,shin17,ahn18,ahn20,ahn21,ahn21b,staros2022combined,ahn2023structural,wines2023quantum}. 
Its full incorporation of electron-electron correlations is not only highly beneficial in describing the vdW interactions, but also enables a simultaneous description of intralayer and interlayer interaction on an equal footing.
Particularly, DMC was able to predict a correct picture of the MIT in bilayer blue phosphorenes with respect to the interlayer separation~\cite{ahn21b}, revealing the band gap reduction mechanism by the interlayer p$_{z}$ hybridization. 
From this, we expect the DMC calculations to provide a deeper and comprehensive understanding of the interlayer coupling in PtSe$_{2}$ beyond the previous DFT studies.

Through DMC calculations, we have investigated the structural stability of various stacking modes in bilayer PtSe$_{2}$. 
For this, we first optimized the monolayer PtSe$_{2}$ structures and then constructed the DMC interlayer binding energy curves in its bilayer forms. Interestingly, our findings reveal that while the AA stacking is most stable, an AB-r stacking is nearly degenerate with the AA stacking within statistical errors.
This stacking degeneracy, which is hardly observed in other TMD materials, underscores a significant role of the interlayer hybridization in offsetting the energy cost associated with a larger interlayer separation in the AB-r mode.
We have utilized our DMC results for bilayer PtSe$_2$ to benchmark the performance of a variety of widely used DFT functionals that are known to represent well vdW binding, interlayer hybridization, or both. From these comparisons, we find that the r$^{2}$SCAN+rVV10 exhibits good performance in describing the interlayer interactions in bilayer PtSe$_2$.
Finally, based on the benchmarked DFT-r$^{2}$SCAN+rVV10 calculations, we have discovered the stacking polymorphism in few-layer PtSe$_{2}$ systems, which turns out to be one of the main factors complicating the characterization of MIT in PtSe$_{2}$.
Our study not only unveils the stacking polymorphism in a few-layer PtSe$_2$ system but also opens new possibilities for nanoelectronic devices based on a refined understanding of its layer-dependent MIT.

\section*{Results and discussion}
\label{sec:results}

\subsection*{Relaxed 1T and 2H monolayer structures}
\begin{figure*}
            \centering
            \includegraphics[width=6.5in]{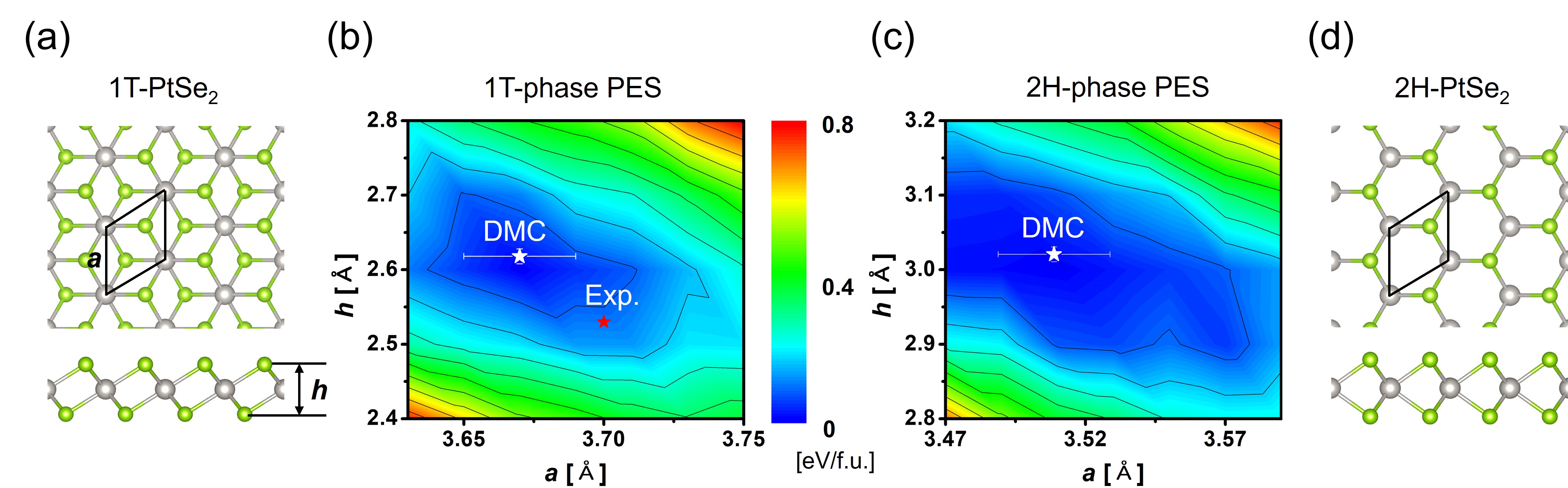}
            \caption{Top and side views of the crystal structures of 1T- and 2H-phase monolayer PtSe$_{2}$ ((a) and (d)), along with their DMC potential energy surfaces (PESs) ((b) and (c)) as functions of lattice constant $a$ and buckling height $h$. The contour plots are in the same color scale denoted by the color table which is in units of eV/f.u.. The white star symbols represent the optimal DMC values of $a$ and $h$ while the red symbol denotes the experimental ones.} 
            \label{fig:mono_DMC}
\end{figure*}

Firstly, we optimized the monolayer structures of 1T and 2H phases using DMC calculations. As depicted in Fig.~\ref{fig:mono_DMC}(a) and (d), the 1T and 2H phases are characterized by an octahedral and a trigonal structure, respectively, each consisting of a transition metal layer sandwiched by two chalcogen layers. Both monolayer phases of PtSe$_{2}$ have been successfully synthesized in recent experiments~\cite{wang2015monolayer,o2016raman,lin2017intrinsically,tong2020phase}. These geometries are defined by lattice parameters and buckling heights labeled as $a$ and $h$ in Fig.~\ref{fig:mono_DMC}(a). We constructed contour maps of DMC potential energy surfaces for varying structural parameters for both 1T and 2H phases, which are presented in Fig~\ref{fig:mono_DMC}(b) and (c). These calculations were conducted using a $4 \times 4 \times 1$ supercell, balancing accuracy and computational cost. The DMC supercell energies were fitted to the 2D Gaussian functions to determine the equilibrium structural parameters, indicated by white stars with statistical errors in Fig~\ref{fig:mono_DMC}(b) and (c). 
Our DMC results for the 1T phase structural parameters align closely with experimental values~\cite{wang2015monolayer}, as illustrated in Fig.~\ref{fig:mono_DMC}(b). 
For the equilibrium monolayer structure of each phase, we made finite-size corrections to the DMC supercell energies by extrapolating to the thermodynamic limit (see Fig. S1 in Supplementary Information~\cite{SI}), which showed an energy difference of 1.922(3) eV/atom between 1T and 2H monolayer phases of PtSe$_2$.

\begin{figure*}
            \centering
            \includegraphics[width=4.5in]{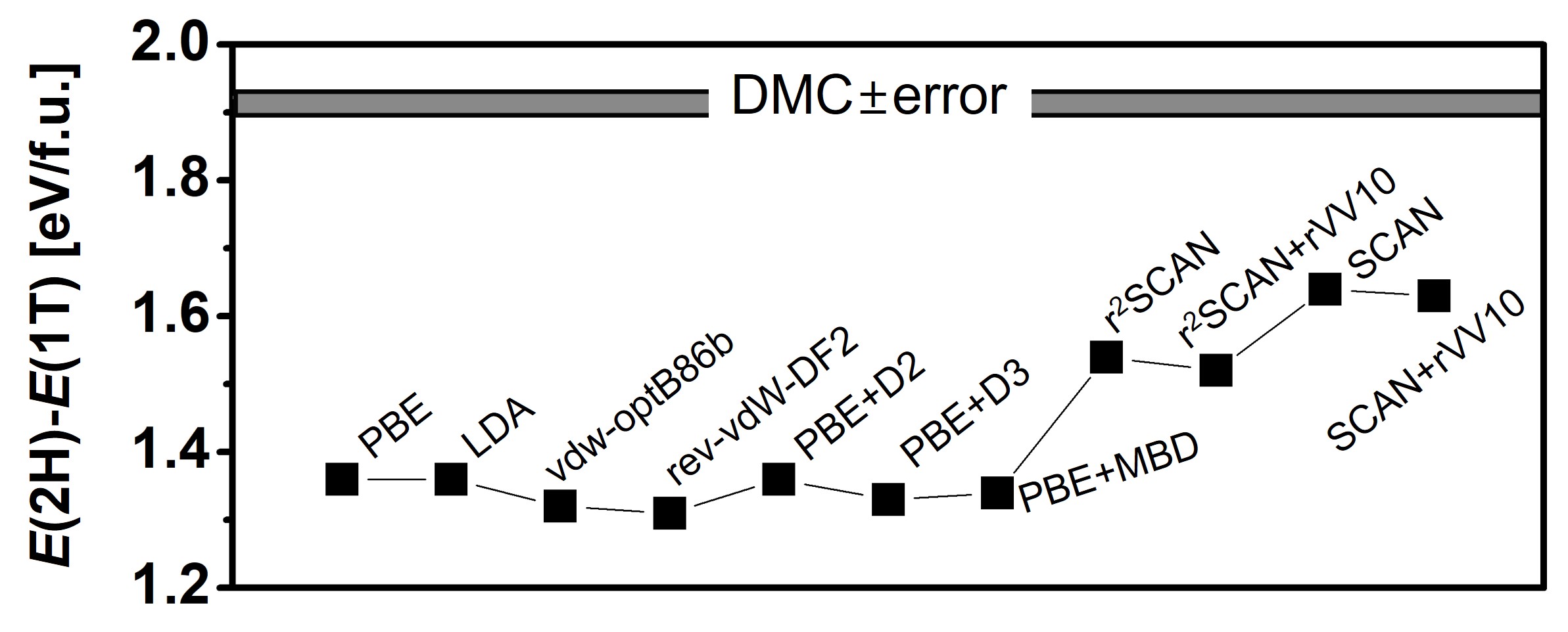}
            \caption{Comparison of the DMC energy difference between 1T- and 2H-phase monolayer PtSe$_{2}$ with the corresponding DFT values computed with several different density functionals.} 
            \label{fig:mono_comparison}
\end{figure*}

In Fig.~\ref{fig:mono_comparison}, we compare this DMC energy difference with corresponding DFT results, where the DMC value is represented by a gray horizontal bar. One can see that the DFT calculations significantly underestimate the energy difference, when compared to the DMC result, by $\sim$ 0.6 eV/f.u at most.
Furthermore, while incorporating vdW interactions has minimal impact on the monolayer energy difference, 
enhancing the description of electronic correlation with meta-GGA functionals brings the DFT results closer to the DMC value.
The substantial energy difference between two monolayer phases of PtSe$_2$ as elucidated by our DMC calculations explains the rarity of the natural occurrence of 2H phase and its synthesis difficulties~\cite{lin2017intrinsically,tong2020phase}. 

\subsection*{Interlayer binding properties of 1T and 2H bilayers}
\begin{figure*}
            \centering
            \includegraphics[width=6.5in]{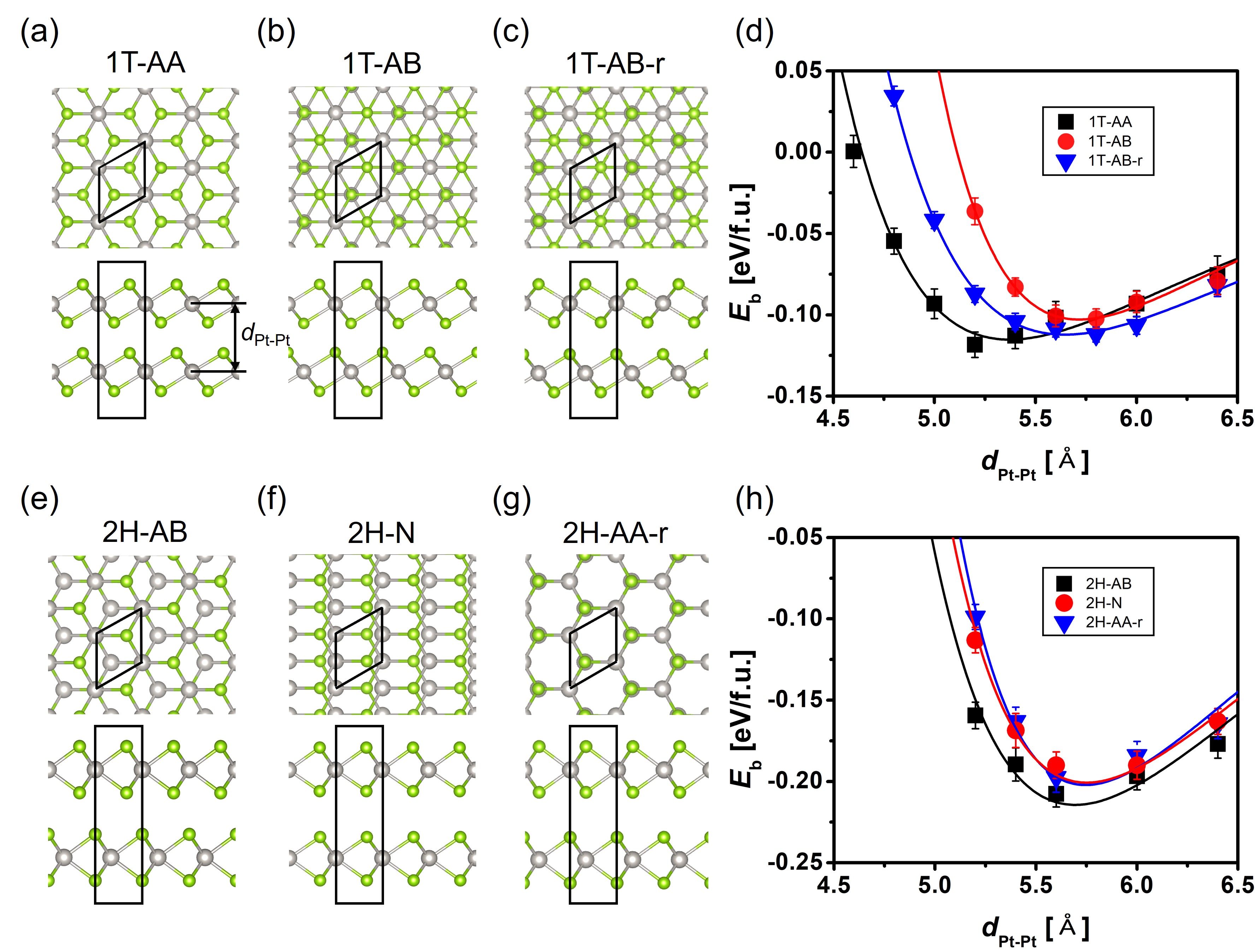}
            \caption{Top and side views of the crystal structure for AA-, AB-, AB-r-stacked bilayer 1T-PtSe$_{2}$ ((a) to (c)) and those for AB-, N-, AA-r-stacked bilayer 2H-PtSe$_{2}$ ((e) to (f)). The black parallelogram represents the unit cell of each structure while $d_{\text{Pt-Pt}}$ in (a) corresponds to the interlayer separation. (d) and (h) present DMC interlayer binding energy curves of bilayer 1T- and 2H-PtSe$_{2}$, respectively, as functions of interlayer distance $d_{\text{Pt-Pt}}$.
            } 
            \label{fig:DMC_binding}
\end{figure*}

\begin{table}
\caption{DMC equilibrium interlayer binding energies (E$_{b}$), interlayer distances ($d_{\text{Pt-Pt}}$), and  total energies relative to the AA stacking ($\Delta$$E_{\text{rel}}$) for 1T- and 2H-phase bilayer PtSe$_2$.}
\begin{center}
\begin{tabular}{cccc}
\hline\hline
~ Stacking ~ & ~E$_{b}$~(eV/f.u.)~ & ~$d_{\text{Pt-Pt}}$~(\AA)~ & ~ $\Delta$$E_{\text{rel}}$~(eV/f.u.) ~\\
\hline
 ~ 1T-AA~ & ~-0.115(3)~ & ~5.37(3)~ & ~ 0.000 ~\\
 ~ 1T-AB~ & ~-0.103(2)~ & ~5.72(2)~ & ~ 0.012(4) ~\\
 ~ 1T-AB-r~ & ~-0.112(2)~ & ~5.64(2)~ & ~ 0.003(4)~\\
 ~ 2H-AB~ & ~-0.214(5)~ & ~5.69(4)~ & ~ 1.823(6) ~\\
 ~ 2H-N~ & -0.201(3)~ & ~5.75(2)~ & ~ 1.836(4) ~\\
 ~ 2H-AA-r~ & ~-0.202(4)~ & ~5.76(2)~ & ~ 1.835(5) ~\\
\hline\hline
\end{tabular}
\end{center}
\label{DMC_bilayer}
\end{table} 

Next, we examined the interlayer binding properties of bilayer PtSe$_{2}$ systems. For this, we constructed DMC interlayer binding energy curves as functions of interlayer distance, d$_{\text{Pt-Pt}}$, for both 1T- and 2H-phase bilayers. Here we considered the three stacking modes for each phase, as shown in Fig.~\ref{fig:DMC_binding}(a)-(c) and Fig.~\ref{fig:DMC_binding}(e)-(g), which were identified experimentally via scanning transmission electron microscopy~\cite{ryu2019situ,xu2021evolution,chen2021atomic}. 
The binding curves, plotted with the finite-size corrected DMC energies (see Fig. S2 in Supplementary Information~\cite{SI}), were fitted to the Morse potentials to determine the equilibrium interlayer binding energies and distances listed in Table~\ref{DMC_bilayer}. 
Figure~\ref{fig:DMC_binding}(d) displays the interlayer binding curves for the 1T bilayer, where the AA stacking mode is shown to be the most energetically stable, mirroring bulk PtSe$_{2}$~\cite{zhao2017high,chen2021atomic}. 
However, our DMC calculations also reveal that another stacking mode of 1T-AB-r, involving a 60$^{\circ}$ rotation of the AB stacking mode, is energetically degenerate to the 1T-AA stacking mode within statistical errors. This 1T-AB-r stacking mode, 
identified to coexist and form grain boundaries with the AA stacking~\cite{chen2021atomic}, has a notably larger interlayer spacing at equilibrium, as presented in the Table~\ref{DMC_bilayer}.
This finding suggests that not only vdW interactions, whose strength depends largely on the interlayer separation, but also significant interlayer hybridization, contributes to this structural degeneracy by offsetting the energy cost associated with larger interlayer distances.
This is further supported by our Bader charge analysis based on the PBE calculations without vdW corrections, 
where the charge transfer in 1T-AB-r stacking (0.014 per Se atom) exceeds that in 1T-AA stacking (0.008 per Se atom).
The observed stacking degeneracy at different interlayer distances may relate to the widely scattered values (5.0~\AA $\sim$ 5.9~\AA) reported for the interlayer distance in bilayer experiments~\cite{shawkat2020thickness,li2021layer,zhang2021precise}.
The increased interlayer distance in the 1T-AB-r stacking mode reduces the spatial overlap between Se-p$_z$ orbitals across the upper and lower layers, leading to diminished splitting into antibonding and bonding states near the Fermi level~\cite{zhao2017high}. This contributes to a larger band gap in the 1T-AB-r mode compared to the 1T-AA stacking mode (see Fig. S3 in Supplementary Information~\cite{SI}).

\begin{figure*}
            \centering
            \includegraphics[width=6.5in]{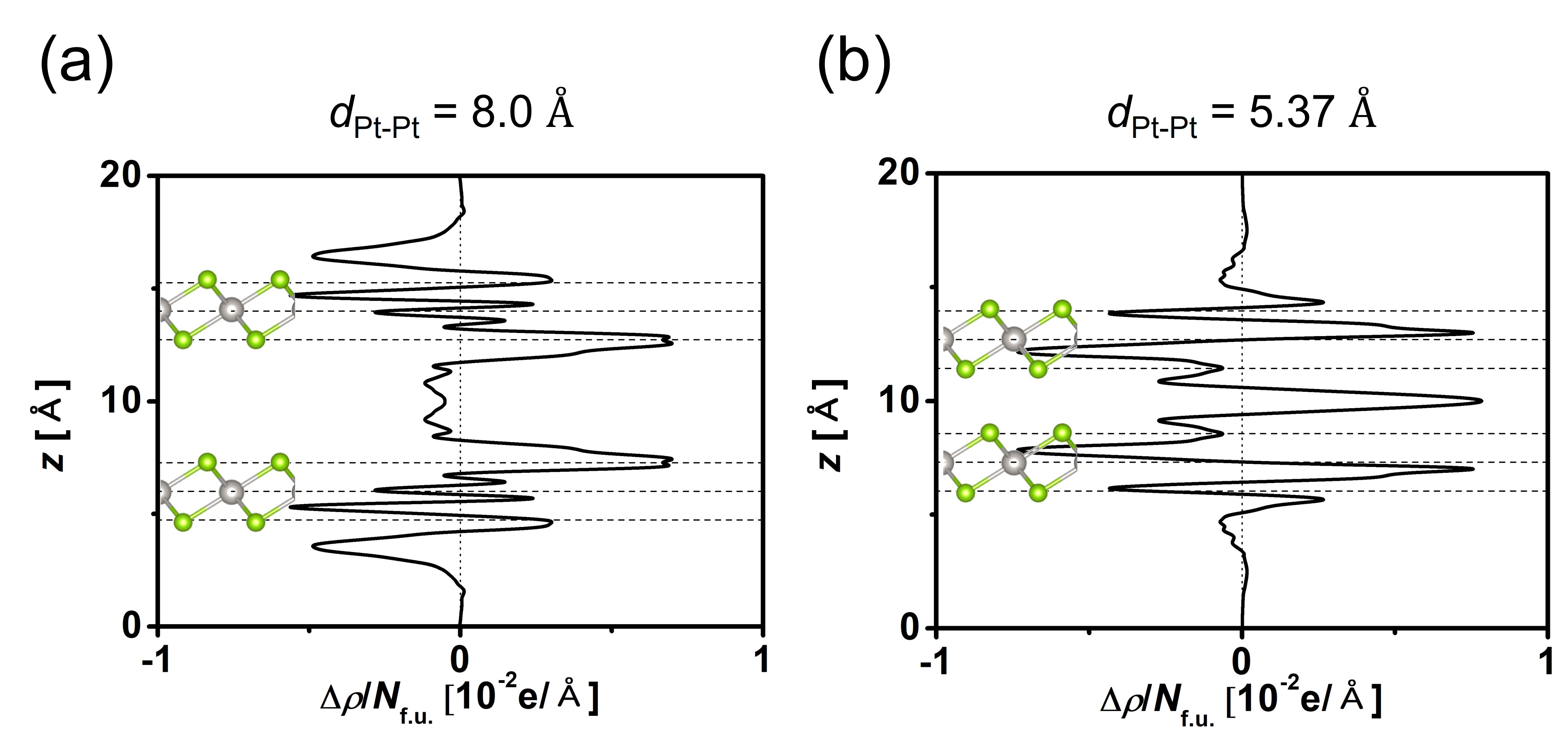}
            \caption{DMC 1D charge density differences per formula unit projected onto the vertical z-axis for the AA-stacked bilayer 1T-PtSe$_{2}$ at interlayer separations of (a) $d_{\text{Pt-Pt}}$ = 8.0~\AA~ and (b) $d_{\text{Pt-Pt}}$ = 5.37~\AA, the latter of which corresponds to the equilibrium interlayer distance. The horizontal dashed lines denote the vertical positions of each Pt and Se planes 
            while the vertical dotted line corresponds to zero density difference.} 
            \label{fig:CDD_comparison}
\end{figure*}

The binding nature of bilayer PtSe$_{2}$ is further analyzed through one-dimensional (1D) charge density profiles projected along the vertical $z$-axis, which allows observation of electron redistribution due to interlayer interactions. For this, we computed the 1D charge density difference between the bilayer and separate monolayers as
$\Delta\rho(z) = \rho^{\text{bi}}(z)-\rho^{\text{upper}}(z)-\rho^{\text{lower}}(z)$ where each DMC charge density is computed using an extrapolated estimator (see Sec. IV Methods). Note that the positive (negative) sign indicates electron accumulation (depletion). Figure~\ref{fig:CDD_comparison}(a) presents the charge density difference profile for an interlayer spacing of 8.0~\AA~in the AA-stacked PtSe$_{2}$ bilayer where the interlayer p$_{z}$ hybridization effects are expected to be minimal and vdW interactions are dominant. 
No charge accumulation is observed in the intermediate region between the upper and the lower monolayers, with slight charge accumulations near the interior Se planes due to intralayer charge redistribution, confirming that interlayer binding is mainly due to vdW interactions driven by fluctuating surface polarization.
On the other hand, Fig.~\ref{fig:CDD_comparison}(b), the charge density difference for the equilibrium distance of $5.37$~\AA, shows significant electron accumulation between the layers due to electron migration from the interior Se planes. This signifies substantial interlayer hybridization to form the covalent bond with electrons shared between the layers, which is sometimes referred as ``quasi interlayer bonding'' between the interior Se planes~\cite{chen2021interlayer,bian2022strong}.
From this, we conclude that the binding nature in PtSe$_{2}$ bilayers is characterized by interlayer hybridization as well as the vdW interactions.

For the 2H-phase bilayer, our DMC results show that the 2H-AB stacking is most energetically favorable, followed by the nearly degenerate 2H-N and 2H-AA-r stackings, with similar equilibrium interlayer separations (see Fig.~\ref{fig:DMC_binding}(e)-(h)).
This energetic ordering between 2H-AB and 2H-AA-r stackings is consistent with the stacking stability observed in other 2H-phase TMD materials such as MoS$_{2}$ and MoSe$_{2}$~\cite{he2014stacking,cortes2018stacking}. Notably, interlayer binding energies for the 2H phase are larger than those for the 1T phase, reducing the energy difference between 1T- and 2H-phase bilayers by about $\sim 0.1$ eV/f.u. (see Table~\ref{DMC_bilayer}), compared to the monolayer energy difference of 1.922(3) eV/atom . 
Our PBE Bader charge analysis shows 
the increased charge transfer in the 2H-AB bilayer (0.020 per Se atom) compared to the 1T-AB-r bilayer, supporting the enhanced interlayer bindings in the 2H phase.

\begin{figure*}[h]
            \centering
            \includegraphics[width=6.5in]{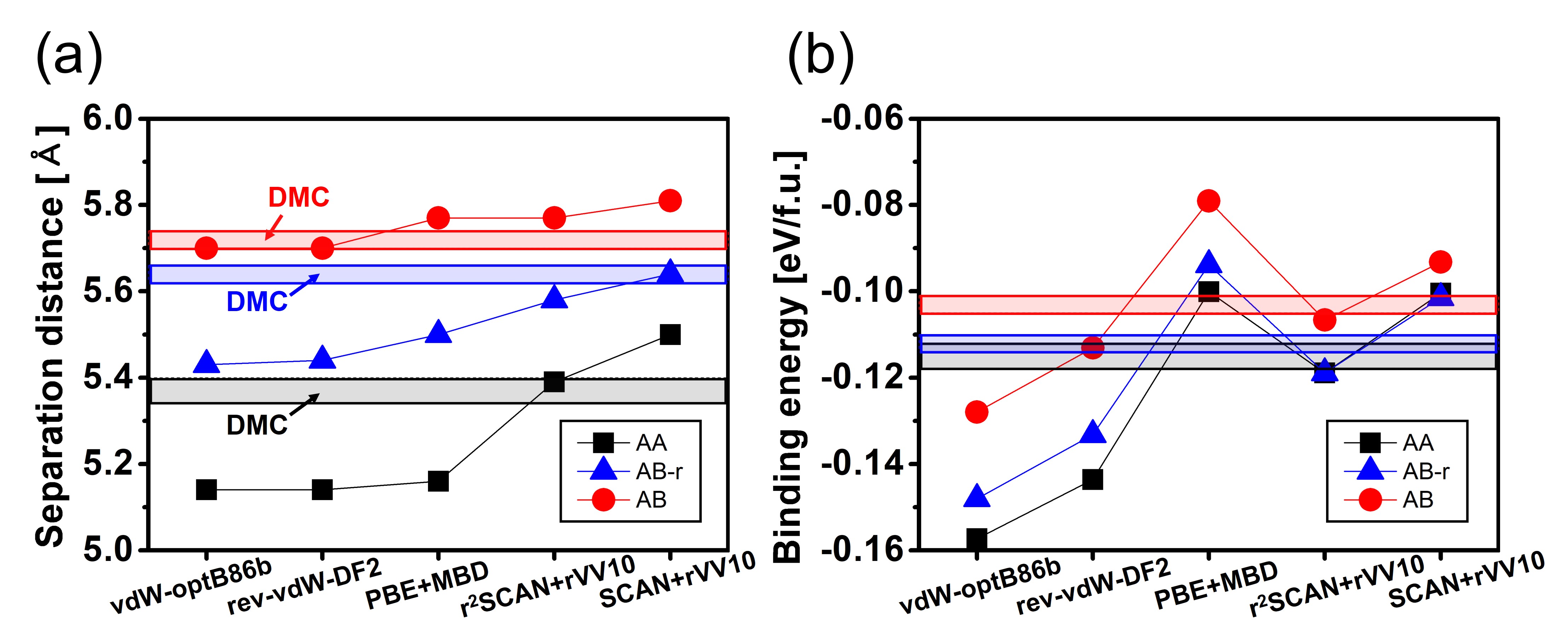}
            \caption{Comparison between DMC and DFT calculations for (a) equilibrium separation distance and (b) interlayer binding energy of AA-, AB-r, and AB-stacked bilayer 1T-PtSe$_{2}$. The DMC optimized values are represented by colored region to reflect their statistical uncertainties.} 
            \label{fig:bi_comparison}
\end{figure*}

We now compare our DMC results for the 1T-phase bilayer PtSe$_2$ with DFT calculations based on various vdW DFT functionals for different stacking modes (AA, AB-r, and AB). Figure~\ref{fig:bi_comparison}(a) and (b) display the DFT equilibrium interlayer distances and binding energies, respectively, with the DMC results represented by colored regions to incorporate their statistical uncertainties. Among the DFT functionsals considered here, the r$^{2}$sCAN+rVV10 functional is shown to produce the most consistent DFT results with the DMC ones for both energetics and separation distances, regardless of the specific stacking mode. The degeneracy between the AA and the AB-r stackings is also corroborated by the r$^{2}$SCAN+rVV10 calculation, underscoring its effectiveness in capturing relative energetics across different bilayer stacking modes. Considering the extreme sensitivity of the  band gap to even small changes in interlayer separation, the geometry optimized at the DFT level with the r$^{2}$SCAN+rVV10 functional is expected to provide a fine reference to assess an accurate band gap of a PtSe$_2$ system.

\subsection*{Stacking polymorphism and band gap}

\begin{figure*}[h]
            \centering
            \includegraphics[width=6.5in]{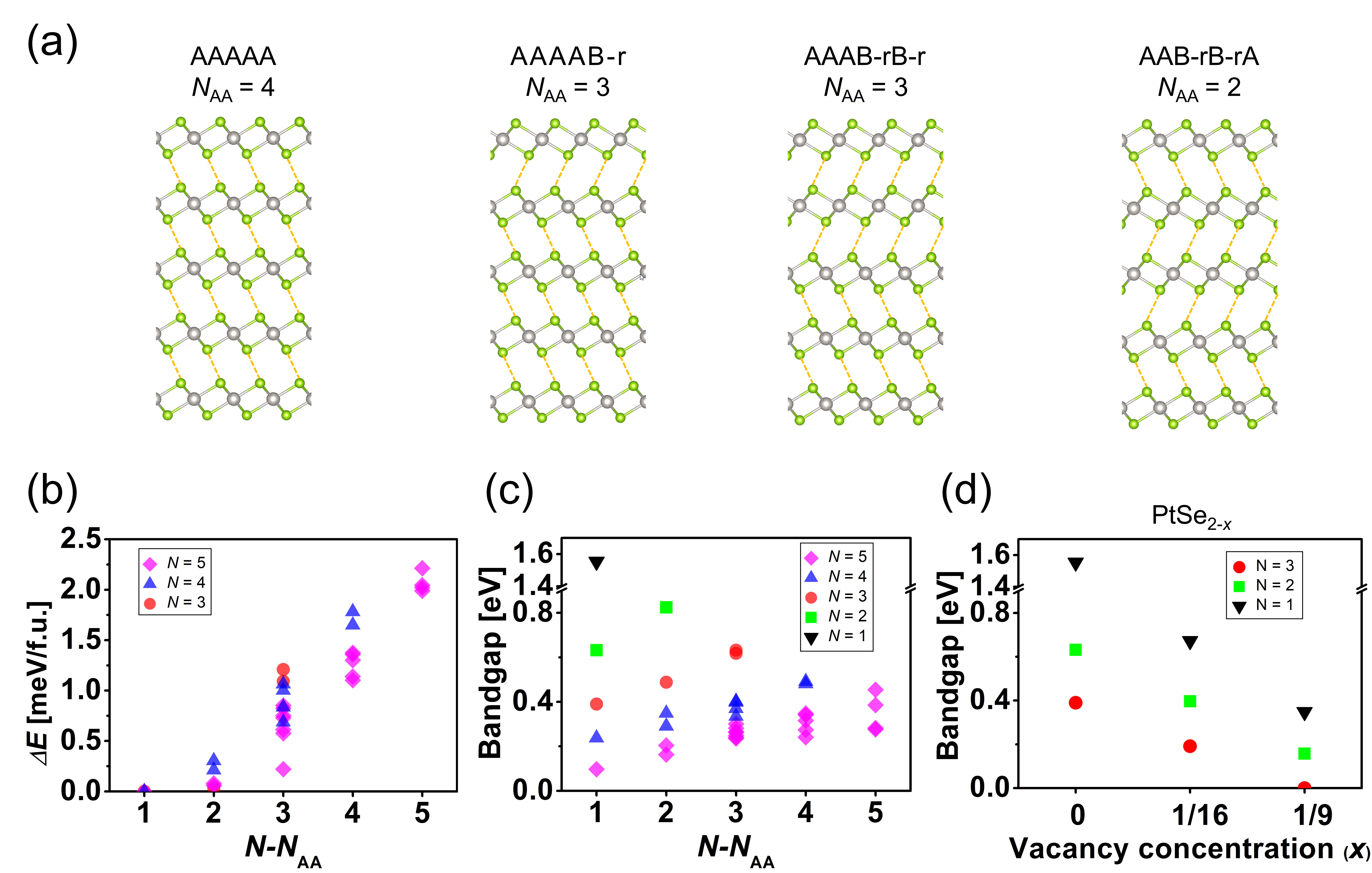}
            \caption{(a) Side views of the crystal structures of AAAAA-, AAAAB-r-, AAAB-rB-r-, and AAB-rB-rA-stacked pentalayer 1T-PtSe$_{2}$ where $N_{\mathrm{AA}}$ represents the number of AA interfaces. (b) presents relative total energies of different $N$-layer stacking sequences of AA and AB-r modes with respect to the stacking sequence composed solely of AA interfaces ($N-N_{\mathrm{AA}}=1$) while (c) shows band gaps of several different polymorphic stacking phases of $N$-layer PtSe$_{2}$. 
            And (d) displays the band gap of AA-stacked  $N$-layer PtSe$_{2-x}$ versus Se vacancy concentration $x$. All data presented here were obtained from benchmarked DFT-r$^{2}$SCAN+rVV10 calculations. 
            }
            \label{fig:stackings_multi}
\end{figure*}

Having verified its accuracy for bilayer systems, we further explored the interlayer binding properties of few-layer 1T-phase PtSe$_2$ using DFT calculations with the r$^{2}$SCAN+rVV10 functional.
Specifically, we assessed the relative energetics of various stacking configurations up to five layers, featuring sequences of AA or AB-r stacking (see Fig.~\ref{fig:stackings_multi}(a) for examples of five-layer stacking). It is noted that these AA and AB-r stackings are nearly energetically indistinguishable for the bilayer.
Figure~\ref{fig:stackings_multi}(b) shows the relative energies of different stacking configurations with respect to the most stable stacking mode, characterized by a successive AA stacking sequence for the given number of layers.
In this figure, one can see several different stacking modes with energies closely matching the lowest energy configuration within a narrow range of 2.5 meV/f.u., indicating a significant stacking polymorphism in few-layer PtSe$_{2}$ systems. This polymorphism is attributed to the near-degeneracy between the AA and AB-r stacking modes observed in bilayer PtSe$_2$. 
Furthermore, as the layer count increases, more stacking configurations fall within this tight energy range, suggesting the potential for stacking fault (disorder) or coexistence of AA with other stacking modes as layer thickness increases.
This stacking polymorphism would lead to grain boundaries between different stacking domains in few-layer PtSe$_{2}$ samples, which could develop metallic electronic structures as discussed in Ref.~\cite{chen2021atomic}.
We also identified similar stacking polymorphism in bulk PtSe$_{2}$ (see Fig. S4 in Supplementary Information~\cite{SI}) as predicted in previous DFT studies~\cite{kempt2022stacking}, influencing the synthesis of polycrystalline PtSe$_2$ samples~\cite{ansari2019quantum,jiang2019large,tharrault2024raman}.
As shown in Fig.~\ref{fig:stackings_multi}(b), the energy difference between an $N$-layer system with only AA stackings ($N-N_{\text{AA}} =1$) and one with only AB-r stackings ($N-N_{\text{AA}} = N$) increases with $N$, reaching $40$ meV/f.u. for bulk PtSe$_2$.
Considering that the AA and AB-r stackings are nearly degenerate in bilayers, this increase suggests that long-range vdW interactions progressively dominate over short-range interlayer hybridization effects with the increasing layer thickness.
A similar polymorphic feature has been reported in other TMD systems, such as MoTe$_{2}$, 
identified experimentally by cross-section visualization of stacking disorder using the scanning transmission electron microscopy~\cite{hart2023emergent},
which calls for analogous measurements for PtSe$_{2}$ systems.

Finally, we discuss the impact of the stacking polymorphism on the band gap of few-layer PtSe$_{2}$. Figure~\ref{fig:stackings_multi}(c) presents the band gaps calculated with the r$^{2}$SCAN+rVV10 functional for various stacking phases of $N$-layer systems. The band gap of the AA-stacked few-layer PtSe$_{2}$ decreases from 1.57 eV in a monolayer to 0.09 eV in five-layer systems. 
However, regardless of the stacking mode, an $N$-layer PtSe$_2$ exhibits a finite band gap for $N \le 5$, indicating no MIT up to five-layer systems.
This is in line with GW calculations of Ansai {\it et~al.}~\cite{ansari2019quantum}, which predicted a semiconducting gap at a four-layer PtSe$_2$ whereas earlier DFT calculations based on semi-local density functionals~\cite{yan2017high,ciarrocchi2018thickness,zhang2021precise,li2021layer} suggested an MIT for $N \le 3$. 
The r$^{2}$SCAN+rVV10 functional thus proves particularly effective at the DFT level for band gap estimations of PtSe$_2$, offering a more cost-efficient alternative  compared to hybrid functionals or GW methods. 
%
However, our DFT-r$^{2}$SCAN+rVV10 results for pristine few-layer PtSe$_2$ do not explain experimental observations of MIT at three to five layers~\cite{yan2017high,ciarrocchi2018thickness,zhang2021precise,li2021layer}. Interestingly, we observed that Se vacancies, the most abundant defects in synthetic PtSe$_{2}$ films~\cite{li2022edge}, significantly reduce the band gap (see Fig.~\ref{fig:stackings_multi}(d)). 
Specifically, 
the band gap of a three-layer PtSe$_2$ with a Se vacancy concentration of $1/9$ disappears entirely, suggesting an MIT at thinner layers than in pristine conditions. 
%
In addition, as seen in Fig.~\ref{fig:stackings_multi}(c), a polymorphic stacking sequence with at least one AB-r interface tends to exhibit less band gap reduction with increasing layer thickness compared to a sequence composed solely of AA interfaces, due to the diminishing band splitting into antibonding and bonding states near the Fermi level with increasing interlayer separation at the AB-r interface.
Therefore, the interplay between stacking polymorphism, potential metallic grain boundaries formed through this polymorphism, and defect formations complicates the MIT in few-layer PtSe$_{2}$, necessitating careful interpretation of its band gap data.

\section*{Conclusion}
\label{sec:conclusion}

This study provides a comprehensive understanding of the interlayer interactions and various stacking configurations of few-layer 1T-phase PtSe$_2$,
employing a combination of DMC and DFT methods.
While the DMC calculations show the near degeneracy between AA and AB-r stacking modes in bilayer PtSe$_2$, our comparative analysis between DMC and DFT results highlights the r$^2$SCAN+rVV10 functional's efficacy in accurate predictions of its structural and electronic properties. 
The DMC-benchmarked DFT study with the r$^{2}$SCAN+rVV10 functional reveals a stacking polymorphism across different layer counts, 
which is likely a contributing factor to the formation of grain boundaries between different stacking domains and the emergence of metallic electronic structures in the few-layer systems.

This work also extends our understanding of the effects of stacking polymorphism and defects on the electronic properties of TMD materials.
Specifically, our discovery of the band gap variability due to the stacking polymorphism, especially the less pronounced band gap reduction in stacking sequences involving AB-r interfaces, suggests that long-range van der Waals interactions and short-range interlayer hybridization collectively influence the electronic properties of PtSe$_2$ as well as its interlayer binding.  
Considering the significant impact of Se vacancies on band gap reduction, the competing influences of vacancy defects and stacking polymorphism   
emerge as crucial factors for analyzing the layer-dependent MIT of few-layer PtSe$_2$. This underscores the complexity of predicting and controlling the electronic behavior of PtSe$_2$ through stacking configuration and layer thickness.

In conclusion, the intricate interplay between stacking polymorphism, defect formations, and interlayer interactions in PtSe$_2$ elucidates the broader challenge of understanding and manipulating the physical properties of TMDs. This study shows that the r$^2$SCAN+rVV10 functional within the DFT framework offers a valuable tool for probing electronic properties of TMD materials as well as their strong interlayer interactions, providing a balance of computational efficiency and accuracy. Future studies should focus on refining the control of stacking faults and defect concentrations to tailor the material properties for specific applications, as well as expanding the investigation to other TMD materials to explore the generalizability of these findings.


\section*{Methods}
\label{sec:methodology}

Our fixed-node DMC calculations were carried out using the QMCPACK package~\cite{kim18,kent20}. We utilized the Slater-Jastrow type trial wave functions, comprising a product of the Slater determinants and correlating Jastrow factors that include up to three-body Jastrow factors to account for electron-electron-ion correlations. The Slater determinants were constructed using Kohn-Sham orbitals from the DFT-PBE calculations executed using the QUANTUM ESPRESSO~\cite{giannozzi09}, with a plane-wave cutoff of 400 Ry and $18 \times 18 \times 1$ Monkhorst-Pack $k$-point mesh~\cite{monkhorst76} for both monolayer and bilayer PtSe$_2$.
We used a norm-conserving scalar-relativistic energy-consistent pseudopotential by Burkatzki, Filippi and Dolg~\cite{burkatzki07,burkatzki08} for Se atoms while Pt atoms were treated with a norm-conserving potential with pseudovalence state of $5s^{2}5p^{6}6s^{1}5d^{9}$, consistent with our earlier DMC studies~\cite{ahn20,ahn2023structural,ahn24}. 
To mitigate one-body size effects, twist-averaged boundary conditions~\cite{lin01} were applied using 36, 16, and 9 twist angles for supercell sizes of $2 \times 2 \times 1$, $3 \times 3 \times 1$, and $4 \times 4 \times 1$, respectively. 
The twist-averaged DMC supercell energies were then extrapolated to the thermodynamic limit ($N = \infty$) through linear fitting to address the two-body size effects. 

The optimization of Jastrow parameters was conducted in  variational Monte Carlo (VMC) calculations with the linear method proposed by Umrigar {\it et al.}~\cite{umrigar07}.
Subsequent DMC calculations were performed with a time step of $\tau = 0.005$ Ha$^{-1}$ and size-consistent T-moves for variational estimation of non-local pseudopotential terms, as in our previous DMC studies for a cousin material of PtTe$_{2}$~\cite{ahn24}.
A vacuum distance was set to be 30~\AA~along the vertical direction to the PtSe$_{2}$ plane to minimize spurious interactions between periodic images.
For the estimation of densities, considering their non-commutative properties with the Hamiltonian, we used an extrapolated estimator defined as $\rho_{\text{ext}} = 2\rho_{\text{DMC}}-\rho_{\text{VMC}}$, 
minimizing trial-function-dependent biases to second order~\cite{foulkes2001quantum}.

Complementary DFT calculations were performed using the VASP package~\cite{Kresse1993,Kresse1996}, where the Kohn-Sham equations were solved with the projector-augmented wave pseudopotentials~\cite{PAW1994,Kresse1999} for both Pt and Se atoms to include spin-obit coupling effects.
A plane-wave cutoff of 400 eV and a $12 \times 12 \times 1$ ($12 \times 12 \times 4$ for bulk) Monkhorst–Pack $k$ mesh were employed.  
The convergence thresholds were set at 10$^{-6}$ eV for electronic states and $5 \times 10^{-3}$ eV/\AA~for forces during structural optimizations.
We employed the meta-GGA functionals of SCAN~\cite{SCAN} and r$^2$SCAN as well as semilocal density functionals of  LDA~\cite{lda} and PBE~\cite{pbe}. These functionals combined with various vdW-corrected schemes, including PBE+MBD~\cite{MBD1}, vdW-optB86b~\cite{vdW-optB88}, rev-vdW-DF2~\cite{rev-vdW-DF2},  SCAN+rVV10~\cite{scan_rVV10}, and r$^{2}$SCAN+rVV10~\cite{r2SCAN_rVV10}, were used to deal with interlayer interactions.

\section*{Data Availability}
The data that support the findings of this study are available
this article, its supplementary material, and the materials
data facility\cite{Blaiszik2016, Blaiszik2019} at [link to be provided upon acceptance of this manuscript].


\begin{acknowledgments}

This work was primarily supported by the U.S. Department of Energy, Office of Science, Basic Energy Sciences, Materials Sciences and Engineering Division, as part of the Computational Materials Sciences Program and Center for Predictive Simulation of Functional Materials. 
J. Ahn (final calculations, writing), H. Shin (editing), A. Benali (editing), and J. T. Krogel (mentorship, analysis, writing) were supported by the U.S. Department of Energy, Office of Science, Basic Energy Sciences, Materials Sciences and Engineering Division, as part of the Computational Materials Sciences Program and Center for Predictive Simulation of Functional Materials. Initial work by J. Ahn was supported by Konkuk University Researcher Fund in 2019. 
Y. Kwon, I. Hong, and G. Lee were supported by the Basic Science Research Program (2018R1D1A1B07042443) through the National Research Foundation of Korea funded by the Ministry of Education. We also acknowledge the support from the Supercomputing Center/Korea Institute of Science and Technology Information with supercomputing resources including technical support (KSC-2020-CRE-0126). 

An award of computer time was provided by the Innovative and Novel Computational Impact on Theory and Experiment (INCITE) program. This research used resources of the Argonne Leadership Computing Facility, which is a DOE Office of Science User Facility supported under contract DE-AC02-06CH11357. This research also used resources of the Oak Ridge Leadership Computing Facility, which is a DOE Office of Science User Facility supported under Contract DE-AC05-00OR22725.

This manuscript has been authored by UT-Battelle, LLC under Contract No. DE-AC05-00OR22725 with the U.S. Department of Energy. The United States Government retains and the publisher, by accepting the article for publication, acknowledges that the United States Government retains a non-exclusive, paid-up,irrevocable, worldwide license to publish or reproduce the published form of this manuscript, or allow others to do so, for United States Government purposes. The Department of Energy will provide public access to these results of federally sponsored research in accordance with the DOE Public Access Plan (http://energy.gov/downloads/doe-public-access-plan).

\end{acknowledgments}

\section*{Reference}
\bibliography{main.bbl}
\end{document}